\documentclass[11pt,a4paper]{article}
\usepackage[nohead, nomarginpar, margin=1in, foot=.25in]{geometry}
\usepackage{graphicx}
\usepackage{color}
\usepackage{amsmath}
\usepackage{amssymb,bm}
\usepackage{comment}
\usepackage[T1]{fontenc}
\usepackage[utf8]{inputenc}
\usepackage{authblk}
\usepackage{csquotes}
\usepackage{subcaption}
\usepackage{soul}
 \usepackage{hyperref}
 \hypersetup{
     colorlinks=true,
     linkcolor=blue,
     filecolor=blue,
     citecolor = blue,      
     urlcolor=cyan,
     hypertexnames=true
     }
\usepackage{verbatim}
\usepackage{url}

\usepackage{cite}
\date{}

\def\@listcomma@comma{\@ifnum{\@tempcnta>\tw@}{,}{}}

\begin{document}
\title{Medium modification type phenomena observed in high-multiplicity p+p collisions at LHC energies}
\author[1]{K. Dey}
\author[2]{A. N. Mishra}
\author[3, \footnote{srikanta.kumar.tripathy@ttk.elte.hu}]{S. K. Tripathy}

\affil[1] {Department of Physics, Bodoland University, Kokrajhar, Assam 783370, India}
\affil[2] {Wigner Research Centre for Physics, 29-33 Konkoly-Thege Mikl\'os str., 1121 Budapest, Hungary}
\affil[3]{Department of Atomic Physics, Eotvos Lorand University, Budapest, H-1117, Hungary}

\maketitle

\begin{abstract}
Nuclear modification factor predicts whether a medium is formed in a collision system or not.  One may verify, if scaling of momentum (transverse) distribution from heavy-ion systems with incoherent superposition of  number of binary collision, is making up the magnitude of elementary like systems. It was observed in a wide range of experimental data that in lower momentum (transverse) region, initial state effects (viz. cold-nuclear matter effect or radial flow) enhances it.  While in higher momentum (transverse) region,  final state effects (viz. jet quenching or re-scattering effects) suppresses it. In this article we made an attempt to predict and explain if this heavy-ion like effect can be seen in elementary collisions at favorable experimental conditions. We have employed PYTHIA8 model in this regard, and used effects which give collective-like behaviour within the model constraint. Results from this model qualitatively explains the behaviour observed in heavy-ion systems.
\end{abstract}


\section{Introduction}\label{intro}
The primary objective of studying hadron-hadron, hadron-nucleus, and nucleus-nucleus collisions at different beam energies is to disentangle the Quark-Gluon Plasma phase from the ordinary hadronic phase of mater and to understand the various aspects of particle production mechanism. The nuclear modification factor $R_{AA}$ or $R_{cp}$  is generally considered as an important tool to study the medium formation in Ultra-relativistic heavy-ion collisions. $R_{AA}$ is defined as the ratio of particle yield in heavy-ion collisions to that of the elementary p+p collisions where both the numerator and denominator are normalized with respect to the average number of binary collisions $\langle \mathrm{N_{bin}} \rangle$ per event in that particular multiplicity class \cite{rcp1}. Similarly, $R_{cp}$ is another way of defining the nuclear modification factor where the reference system is taken as the peripheral collisions in place of elementary $p+p$ collisions.  The definition of the quantity ensures it to be unity in absence of any medium effect. The suppression (i.e. $R_{cp}$ or $R_{AA}<1$) of high $p_{T}$ particles is believed to be caused by either the partonic energy loss suffered while traversing through the hot and dense medium or the cold-nuclear matter effect (CNME). On the other hand, the rescattering driven final state phenomenon called the Cronin effect might also be responsible for $R_{cp}$ or $R_{AA}$ to be greater than unity (enhancement) at the intermediate $p_{T}$ region \cite{cronin_1, cronin_alt}. The magnitude of $R_{cp}$ or $R_{AA}$ may also be influenced by the initial state effects such as nuclear shadowing \cite{shadowing1, shadowing2}. The nuclear modification factor has been extensively studied for different colliding systems at various experimental heavy-ion facilities such as SPS \cite{sps1, sps2}, RHIC \cite{rhic1, rhic2, rcp1, rhic4}, and LHC \cite{lhc1, lhc2, lhc3} at different beam energies. The observation of large suppression ($R_{cp}$ or $R_{AA}<1$) of high $p_{\mathrm{T}}$ hadrons in central Au+Au/Pb+Pb collisions while no suppression for electroweak boson ($Z^0$) \cite{cms_wz, atlas_wz} and direct photons \cite{direct_photon1, direct_photon2} in Pb+Pb and p+Pb collisions provides the strong evidence that the produced hot and dense (partonic) medium is responsible for the observed suppression pattern\footnote{$Z^0$ and direct $\gamma$ donot possess color charge and thus not effected by the presence of strongly interacting medium.}. 
 In order to investigate the onset of medium formation, a study was carried out by the STAR collaboration during the RHIC-BES program \cite{bes1}. In that study, $R_{cp}(p_T)$ was calculated for a wide range of beam energies such as $\sqrt{s}=$ 7.7, 11.5, 19.6, 27, 39, 200 GeV. As expected, the magnitude of $R_{cp}$ was seen to increase with the decrease of beam energy and becomes greater than unity below $\sqrt{s}$ < 39 GeV. This result does not necessarily rule out the presence of partonic energy loss below $\sqrt{s}$ < 39 GeV. This is because of the fact that at lower energies the Cronin like enhancement competes with the suppression effect caused by the partonic energy loss \cite{bes3}. 

On the other extreme, the study of multiparticle production in high-multiplicity (HM) $p+p$ collision is getting considerable attention recently due to its startling resemblance with heavy-ion collision. The recent results on strangeness enhancement in HM $p+p$ collisions measured by the ALICE collaboration \cite{nature_pp} and, in particular, the observation of  strangeness ordering is found to be precisely in line with the results obtained in various heavy-ion collision experiments at SPS, RHIC, and LHC. It may further be added that the CMS and ATLAS collaboration at LHC has also observed a long-range correlation i.e., ridge-like structure in the two-particle azimuthal correlation in HM p+p collisions, which has a striking similarity with that of heavy-ion collision \cite{ridge_pp1, ridge_pp2, ridge_pp3, ridge_pp4}.  The origin of such heavy-ion-like behavior in HM $p+p$ collisions is a matter of debate and one may ask the following questions: Whether QGP-droplets were produced in the violent collisions in small systems? or the apparent collectivity is an artefact of the result of an extended network of interactions among the partons? The color reconnection (CR) model within the PYTHIA8 framework could able to explain the flow-like collective behavior \cite{CR-1} while the combination of rope hadronization mechanism and QCD-based color reconnection successfully explained the strangeness enhancement pattern observed by the ALICE collaboration at LHC \cite{strange_RH}.
Although, different observables were studied both theoretically and experimentally, no study on nuclear modification type phenomena for HM $p+p$ collisions was undertaken. It would be very much interesting to see if at all any heavy-ion-like behavior could be seen in HM $p+p$ collisions in the context of medium modification factor. In the present investigation, an attempt has been made to study the medium modification factor in HM $p+p$ collisions with the help of PYTHIA8 model.\\
The paper is organized in the following way. In section \ref{pythia}, a detailed account regarding the PYTHIA8 model is presented. 
Section \ref{R_and_D} is devoted to analysis techniques, results and discussions. Finally, in section \ref{summary} we summarized our results.

\section{The PYTHIA model}
\label{pythia}
PYTHIA ~\cite{pythia6,pythia8.1,pythia8.2} is a general-purpose Monte Carlo event generator used for the generation of events in high-energy collisions of leptons, protons, and nuclei. 
The event generation in PYTHIA8 consists of several steps starting typically from a hard  scattering  process, followed by initial and final state  parton showering, multiparton interactions, and the final hadronization  process. Two processes in PYTHIA8 generate collective phenomena like behaviour, namely MPI (Multiple Parton Interaction) and CR (Color Reconnection) which are briefly discussed below. 

\subsection{Inclusion of MPI in PYTHIA8} 
The high-energy hadronic collisions are expected to have multiple parton-parton interactions in the same events because the projectile and the target beam-particles contain a multitude of partons that can interact~\cite{TS:1987}. The inclusion of MPI in PYTHIA8~\cite{pythia6} is an important assumption and has been supported by many experimental results~\cite{Akesson:1987,Alitti:1991,Abe:1997}. The MPI allows to have a qualitatively good description of the multiplicity distributions as well as the correlation of observable like transverse spherocity with multiplicity in $p+p$ collisions at the LHC energies~\cite{aliceS0:2012,aliceS0:2019}. 

The differential cross-section for patron-patron scattering in PYTHIA8 may be written as \cite{nmpi_1}:
\begin{equation}
    \frac{d\sigma}{dp_\mathrm{T}^2} = \sum_{i,j,k} \iiint f_i \left( x_1,Q^2 \right) f_j \left(x_2,Q^2 \right) \frac{d\hat{\sigma}_{ij}^k}{d\hat{t}} \delta \left ( p_\mathrm{T}^2 - \frac{\hat{t}\hat{u}}{\hat{s}}\right ) dx_1 dx_2 d\hat{t}
\end{equation}
Where $f_i$ are structure functions for finding patron $i$; $s$, $t$ and $u$ are Mandelstream variables and $Q^2$ used as $p_\mathrm{T}^2$ factorisation.
In PYTHIA8, average of inelastic non-diffractive events shown to be related with nMPI follows:
\begin{equation}
    \langle \mathrm{nMPI}(p_{\mathrm{T}_{min}}) \rangle = \frac{1}{\sigma_{nd}} \int_{p_{\mathrm{T}_{min}}^2}^{s/4} \frac{d\sigma}{dp_\mathrm{T}^2}dp_\mathrm{T}^2
\end{equation}
Where $\sigma_{nd}$ is inelastic non-diffractive events cross-section, the integral represents integrated cross-section on a specific $p_{\mathrm{T}_{min}}$ scale.
Value of above equation higher than unity corresponds more than one such sub-collision per event.

Generation of consecutive MPIs is formulated in PYTHIA8, as an evolution downwards in $p_\mathrm{T}$. Correction of the probability distribution of $p_\mathrm{T}$ is done by an exponential factor, so that for $p_{\mathrm{T}_1}$ to be hardest interaction, there must not be any interaction range between $\sqrt{s}/2$ and $p_{\mathrm{T}_1}$.
\begin{equation}
    \frac{d\mathrm{P}}{dp_{\mathrm{T}_i}} = \frac{1}{\sigma_{nd}} \frac{d\sigma}{dp_{\mathrm{T}_i}} exp \left ( - \int_{p_{\mathrm{T}_i}}^{p_{\mathrm{T}_{i-1}}} \frac{1}{\sigma_{nd}} \frac{d\sigma}{dp_ \mathrm{T}^\prime} dp_\mathrm{T}^\prime \right )
\end{equation}

\subsection{Inclusion of CR in PYTHIA8} 
We have used MPI-based original PYTHIA8 scheme in this work \cite{cr_1}. The hadronisation mechanism included in PYTHIA8 is CR. 
This is a microscopic mechanism in which final partons are connected by color string, in such a way that the total string length becomes as short as possible~\cite{pythia6}.  One string connecting two partons follows the movement of the partonic end points.  The effect of this movement is a common boost of the string fragments (hadrons).  With CR two partons from independent hard scattering at mid-rapidity can color reconnect and make a large transverse boost. This effect becomes more important in single events with MPI. This boost effect caused by CR is similar to how flow affects hadrons in hydrodynamics, but the origin of the boost is clearly different in CR compared to hydrodynamics~\cite{CR-1}.

With hardness scale $p_T$, reconnection probability for an interaction may be written as:
\begin{equation}
    P_{rec} (p_T) = \frac{(R_{rec} \times p_{T0})^2}{(R_{rec} \times p_{T0})^2+p_{T}^2 }
\end{equation}
Where $R_{rec}$ is the CR strength (0-10) and $p_{T0}$ is the $p_T$ regularisation parameter (energy dependent parameter used to damp the low $p_T$ divergence). 

Here, an interaction tries to reconnect with next-highest one in $p_T$. If this is not satisfied then consecutively higher ones are tried. This makes sure that the total reconnection probability for an interaction is $1 - (1 - P_{rec})n$, n being number of interactions at higher $p_T$ scales. 

The reconnection probability chosen to be higher for soft systems. Here all gluons of lower $p_T$ interactions were inserted onto the colour flow dipoles of a higher- $p_T$ ones. This also makes sure to minimise the total string length ($\lambda$) \cite{pythia8.2}. This may be written as:
\begin{equation}
    \Delta \lambda = ln \frac{(p_i p_k)(p_j p_k)}{(p_i p_j)m_0^2},
\end{equation}
 where $m_0$ is a typical hadronic mass scale.

The results reported in this paper are obtained from 300 Million non-diffractive events for $p+p$ collisions at at $\sqrt{s}=$ 200 GeV, 900 GeV, 5.02 TeV and 13 TeV, simulated using PYTHIA version 8.240 with the default Monash 2013 tune~\cite{pythiaMonash}.

\section{Results and discussions}\label{R_and_D}
We have considered the charged hadrons ($\pi^\pm$, $k^\pm$, $p$ and $\bar{p}$) within the ALICE detector acceptance i.e. $|\eta| < 0.8 $ for this analysis.
In order to show the robustness of the present study, two different multiplicity class selection criteria are adopted. In the first method we used the number of semi-hard partonic scattering per event (or number of multiple partonic interaction per event) \cite{nmpi} as a measure of the multiplicity variable. In the second method, multiplicity class is estimated by measuring charge particle multiplicity at mid-rapidity window ($|\eta|< 0.8$).

 \begin{figure}[htbp]
   \centering
   \includegraphics[scale=0.45]{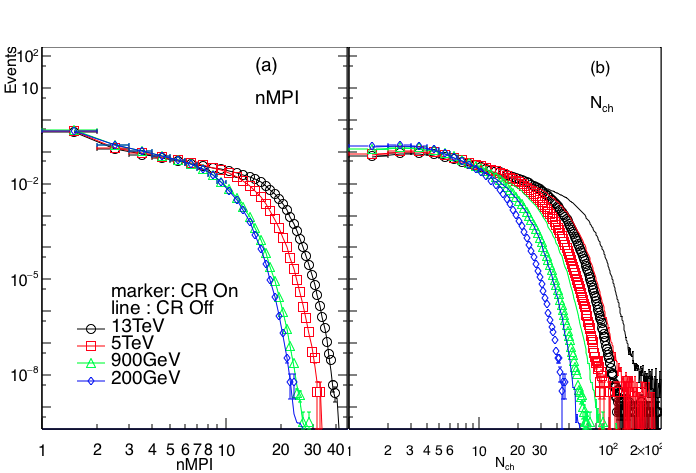}
   \caption{(Color online) nMPI and $N_{ch}$ distributions ($|\eta|<0.8$). Markers represent default PYTHIA8 (i.e. CR = on), while lines denote CR=off.}
    \label{fig:nCh}
    \end{figure}
    
Left side of Fig. \ref{fig:nCh} shows  nMPI distribution, while right side shows $N_{ch}$ distribution, for all energies. Results from default PYTHIA8 (i.e. CR = on) is shown in markers, while solid lines denote CR=off. As expected, we observe a clear energy dependent nMPI and $N_{ch}$ values, 13 TeV being highest while 200 GeV being lowest.
Also we observe that nMPI values are independent of CR, giving same numbers interactions. 
From right side of Fig. \ref{fig:nCh}, we see for higher multiplicity window, number of charged particles (in mid-rapidity) is higher while CR is off. This is consistent with the previous observation \cite{cr:2016} that at very low multiplicity, effect of CR is very weak (for events with less than half of the mean multiplicity). Also we observe this difference is higher, as the energy increases from $\sqrt{s}$ = 200 GeV to $\sqrt{s}$ = 13 TeV.  More percentage of charged particles are produced in $\sqrt{s}$ = 13 TeV while CR is off, than the case in $\sqrt{s}$ = 200 GeV.

Several studies were carried out at RHIC (STAR \cite{ppmult_1} at  $\sqrt{s}=$ 200 GeV) and LHC (ALICE, CMS \cite{ppmult_2}) where multiplicity classes were determined based on the charged particle multiplicity $N_{ch}$ measured at mid-rapidity. However, the cuts on $N_{ch}$ based on which multiplicity were determined seems arbitrary and this changes from experiment and collision energy.
In the present study we therefore have formulated a novel method to determine high-multiplicity and low-multiplicity class. Our method is based on the Lattice QCD expectation of the formation of a non-hadronic medium beyond the critical energy density of 1 $\mathrm{GeV/fm^{3}}$. According to Bjorken, this 1 $\mathrm{GeV/fm^{3}}$ energy density limit corresponds to the $\mathrm{d}N_{ch}/ \mathrm{d}\eta = 12$. Based on accelerating hydrodynamic description, an improved initial energy density estimation of the QGP were made by  M\'at\'e Csan\'ad et al. \cite{Mate_enDen, Mate2} according to which the charge particle multiplicity corresponds to critical energy density is $\mathrm{d}N_{ch}/ \mathrm{d}\eta = 9$. In the present study, the event class for which $\mathrm{d}N_{ch}/ \mathrm{d}\eta > 9$ are therefore considered as the most central or High-Multiplicity class.

Alternatively, one can argue of using multiplicity class based on percentage of multiplicity distribution (nMPI or $N_{ch}$) as followed in heavy-ion collisions, however this will add ambiguity to multiplicity class definition. For example,  0-5\% multiplicity class of 13 TeV have that nMPI value, which $\sqrt{s}$ = 200 GeV may not reach at all. In this case, one will compare physics results, for highest multiplicity class of one energy with, with possibly mid/lowest multiplicity class of another energy.

\begin{figure}[htbp]
\centering
\includegraphics[scale=0.45]{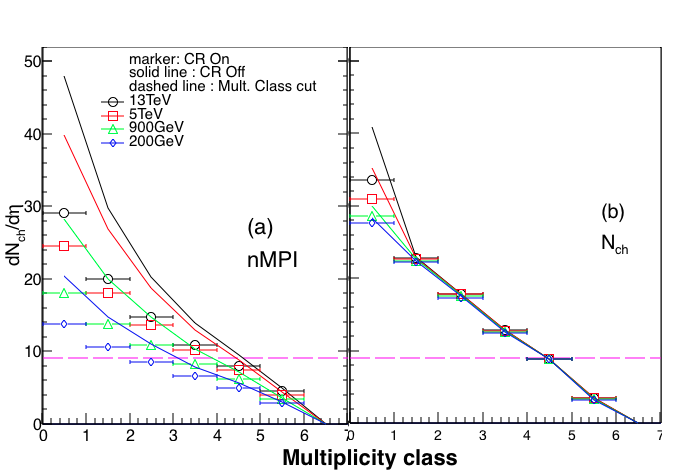}
\caption{(Color online) $dN_{ch}/d\eta$ distributions from nMPI and $N_{ch}$ multiplicity class definition. Markers represent CR on, while solid lines representation as CR off. The dashed line represents the cut off value ($dN_{ch}/d\eta$ >9) for HM class.}
\label{fig:dNchdEta}
\end{figure}

In Fig. \ref{fig:dNchdEta} we have shown $dN_{ch}/d\eta$ distribution for the produced charged particles for $\sqrt{s}=$ 200 GeV, 900 GeV, 5.02 TeV and 13 TeV. Results from default PYTHIA8 (i.e. CR = on) is shown in markers, while solid lines denote CR=off. 
When multiplicity class definition is based on same number of $N_{ch}$ for all energies, it should give similar values of $dN_{ch}/d\eta$ except for multiplicity class 0 (highest multiplicity class). As for highest multiplicity class, upper bound of $N_{ch}$ is  different for different energies. So only in this multiplicity class, results are different. However, this should differ for nMPI, as similar number of nMPI, might correspond different number of particle produced in different energies.

We observe, for all energy, multiplicity class estimated from nMPI, $dN_{ch}/d\eta$ > 9 for multiplicity class 0 and 1. So these two classes may be used as high multiplicity classes. For multiplicity class estimated from $N_{ch}$, first four classes may be taken as high multiplicity class, where $dN_{ch}/d\eta$ >9. So we took class 1 (instead of class 0 for statistical reasons) as HM class, and class 5 as LM class, when multiplicity class estimated from nMPI. To make $N_{ch}$ results consistent with nMPI, we took same class definition for $N_{ch}$ as well.

In heavy-ion collisions for the calculation of nuclear modification factor, $p_T$ spectra are scaled by incoherent nucleon-nucleon collision ($N_{coll}$ or $N_{bin}$ used interchangeably for $R_{AA}$ or $R_{cp}$). In $p+p$ collisions, this scale factor should be unity. But the observation of collective effects in HM  $p+p$ collisions compelled us to explore the other possibilities of calculation of this scale factor.  As reported in ref. \cite{pColl_1,pColl_2}, average number of pions $\langle N_\pi  \rangle$ can also be used as a measure of multiplicity variable. They argued that the average number of participating nucleons $\langle N_{part} \rangle$, which is generally considered as multiplicity variable in AA/pA collisons, exhibits a nonlinear behavior with the volume of the participant zone, while $\langle N_\pi  \rangle$ shows perfect participant scaling. In this present investigation, we have therefore explored the average number of pions $\langle N_\pi  \rangle$ as the scaling factor.

Tab. \ref{Tab:cent} listed average number of pions along with $dN_{ch}/d\eta$ values for $\sqrt{s}=$ 200 GeV, 900 GeV, 5.02 TeV and 13 TeV for CR on. When we use multiplicity class definition from $N_{ch}$, $dN_{ch}/d\eta$ and average number of pions (both from from mid-eta range) do not make the much difference with increase of energy ( 65 fold, from 200 GeV to 13 TeV). However when we use multiplicity definition from nMPI, we see energy dependence is visible.

\begin{table}
\centering
\caption{The values of $\langle dN_{ch}/d\eta \rangle$ and $\langle N_{\pi} \rangle$ for CR on.}
\begin{tabular}{ p{3cm} p{2cm} p{1.5cm} p{1.5cm}  p{1.5cm} p{1.5cm}  p{1.5cm}}
 \hline
 $\sqrt{s}=13$\,TeV &nMPI &$\langle dN_{ch}/d\eta \rangle$ &$\langle N_{\pi} \rangle$ &$N_{ch}$ &$\langle dN_{ch}/d\eta \rangle$ &$\langle N_{\pi} \rangle$\\
 \hline
Class 1 &7-11 &19.8766 &17.1302 &20-25  &22.8405 &19.6821 \\
Class 5 &0-2 &4.41779 &3.86645 &0-7  &3.54135 &3.09736 \\
  \hline
 $\sqrt{s}=5$\,TeV &nMPI &$\langle dN_{ch}/d\eta \rangle$ &$\langle N_{\pi} \rangle$ &$N_{ch}$ &$\langle dN_{ch}/d\eta \rangle$ &$\langle N_{\pi} \rangle$\\
 \hline
Class 1 &7-11 &18.0231 &15.5655 &20-25  &22.7337 &19.6617 \\
Class 5 &0-2 &3.98755 &3.49917 &0-7  &3.49725 &3.06125 \\
 \hline
 $\sqrt{s}=900$\,GeV &nMPI &$\langle dN_{ch}/d\eta \rangle$ &$\langle N_{\pi} \rangle$ &$N_{ch}$ &$\langle dN_{ch}/d\eta \rangle$ &$\langle N_{\pi} \rangle$\\
 \hline
Class 1 &7-11 &13.7882 &11.9754 &20-25  &22.448 &19.5901 \\
Class 5 &0-2 &3.35662 &2.95904 &0-7  &3.33711 &2.92919 \\
 \hline
 $\sqrt{s}=200$\,GeV &nMPI &$\langle dN_{ch}/d\eta \rangle$ &$\langle N_{\pi} \rangle$ &$N_{ch}$ &$\langle dN_{ch}/d\eta \rangle$ &$\langle N_{\pi} \rangle$\\
 \hline
Class 1 &7-11 &10.5998 &9.25801 &20-25  &22.1894 &19.5339 \\
Class 5 &0-2 &2.86502 &2.53744 &0-7  &3.15316 &2.77724 \\
 \hline
\end{tabular}
\label{Tab:cent}
\end{table}

\subsection{Transverse momentum distribution}\label{pt-dist}
\begin{figure}[htbp]
\centering
\includegraphics[scale=0.45]{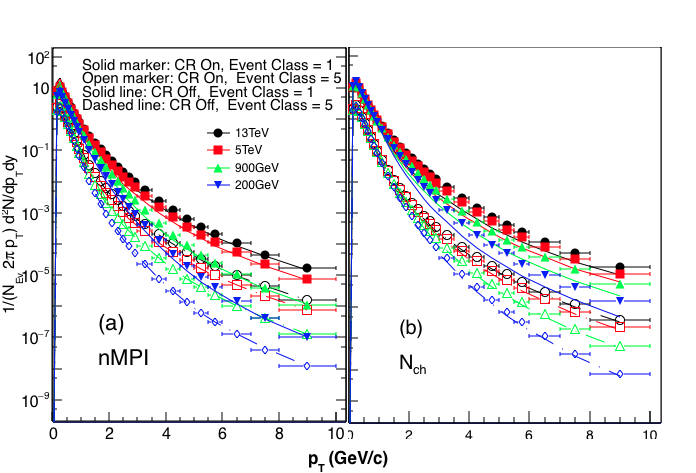}
 \caption{(Color online) Transverse momentum distribution ($p_\mathrm{T}$) of the produced charged particles ($\pi^{\pm}$, $k^{\pm}$, $p$, $\bar{p}$) at the studied energies for both high and low multiplicity classes. The figure on the left (a) and right side (b) corresponds to the multiplicity definition by means of nMPI and $N_{ch}$ respectively. Markers represent CR is on (solid marker for class-1 and open for class-5), while lines represent as CR off (solid line for class-1 and dashed line for class-5).}
 \label{fig:pTSpectra}
\end{figure}

Transverse momentum distribution of the produced charged particles ($\pi^{\pm}$, $k^{\pm}$, $p$, $\bar{p}$) at $\sqrt{s}=$ 200 GeV, 900 GeV, 5.02 TeV and 13 TeV are plotted in Fig. \ref{fig:pTSpectra} for low and high multiplicity classes using two method of multiplicity determination in PYTHIA8 with and without the CR mechanism. The solid and the dashed lines respectively represents the CR off for multiplicity class 1 and class 5 for respectively. Similarly the solid markers and the open markers corresponds to the multiplicity class 1 and class 5 respectively. Different colors represent different beam energies.

The multiplicity dependence of the $p_\mathrm{T}$ spectra is clearly visible for both the multiplicity determination methods. As expected the $p_\mathrm{T}$-spectra becomes harder for high-multiplicity classes than their low-multiplicity counterparts at all the studied energies. We also observe that the CR scenario results higher magnitude of $p_T$ spectra in all cases and this difference is most visible for HM classes. For multiplicity class from $N_{ch}$, difference between CR on to CR off, is significant in central class than peripheral class.

\subsection{Medium modification factor}
Inspired by the definition of nuclear modification factor in nucleus-nucleus (AA) collision, we have redefined the nuclear modification factor for elementary p+p collisions in the following way

\begin{align}
p_{cp} = \frac{ [d^2N/dp_{\mathrm{T}}dy / \langle N_x \rangle  ]^{\text{cent}}} { [d^2N/dp_{\mathrm{T}}dy /\langle N_x \rangle]^{\text{peripheral}}}
 \end{align}

and call it as medium modification factor. Where the normalization factor $\langle N_x \rangle$ represents the number of pions per event within $|\eta|<0.8$.

\begin{figure}[htbp]
\centering
\includegraphics[scale=0.45]{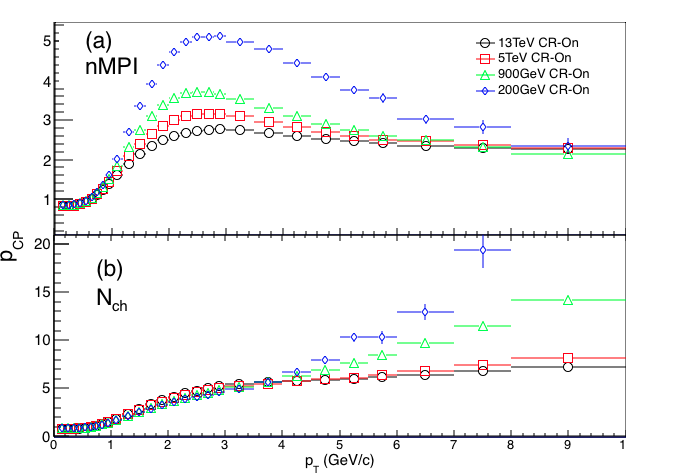}
 \caption{(Color online) Medium modification factor $P_{cp}$ as a function of $p_\mathrm{T}$. In the top (a) and the bottom panel (b) the multiplicity classes are computed from from nMPI and $N_{ch}$ (within $|\eta|<0.8$) respectively.}
 \label{fig:pCP_CRon}
\end{figure}

In Fig. \ref{fig:pCP_CRon}, the medium modification factor $P_{cp}$ for charged particles are plotted as a function of transverse momentum at $\sqrt{s}=$ 200 GeV, 900 GeV, 5.02 TeV and 13 TeV $p+p$ collisions for both of the multiplicity determination method using PYTHIA8. Fig.\ref{fig:pCP_CRon}(a) represents the $P_{cp}$ when multiplicity class is determined by means of nMPI. It can be clearly seen that the value of $P_{cp}$ increases as a function of $p_\mathrm{T}$, reaches a maximum at the intermediate $p_\mathrm{T}$ and then decreases towards higher $p_\mathrm{T}$ values for all the studied energies. The magnitude of $P_{cp}$ at the intermediate $p_{\mathrm{T}}$ is seen to be smaller for higher energies. This qualitative behavior resembles the trend experimentally observed for heavy-ion collisions at the RHIC BES program \cite{bes1}. Although suppression was observed in RHIC BES beyond $\sqrt{s}=$ 39 GeV \cite{bes1}, no suppression is seen for HM $p+p$ collisions at all the energies considered for the present study up to $\sqrt{s}=13$ TeV. At high $p_\mathrm{T}$, the $P_{cp}$ is observed to be independent of the beam energy. Similarly $P_{cp}$ is shown for the  multiplicity class estimator $N_{ch}$ in figure \ref{fig:pCP_CRon}(b). 
One can see an overall monotonic increase of $P_{cp}$ values as a function of $p_\mathrm{T}$ for all energies. It may further be noticed from the figure that the $P_{cp}$ values are independent of energy up to a transverse momentum of roughly 4 GeV/c, while beyond 4 GeV/c, $P_{cp}$ values are observed to be larger for lower energies. The multiplicity class selection method based on $N_{ch}$ is believed to be associated with multiplicity class bias or auto-correlations, since the both of particle selection and multiplicity class selections are done by taking the same phase-space or $\eta$-window i.e. $|\eta|<0.8$.

Even after adopting two ways of defining the multiplicity class, we get $P_{cp}$ to be higher than unity for all energies. This findings (within the model formalism constraint) tells us that we do not see any medium effect as such in $p+p$ collisions from $\sqrt{s}$ = 200 GeV to $\sqrt{s}$ = 13 TeV. However we observe $R_{cp}$ like behavior as seen for heavy-ion collisions in these elementary $p+p$ collisions. In order get more insight let us now discuss the possible final state effects which might cause this type of behavior in elementary collisions.

\subsection{Effect of Color Reconnection}\label{effect_CR} 
\begin{figure}[htbp]
\centering
\includegraphics[width=0.7\linewidth]{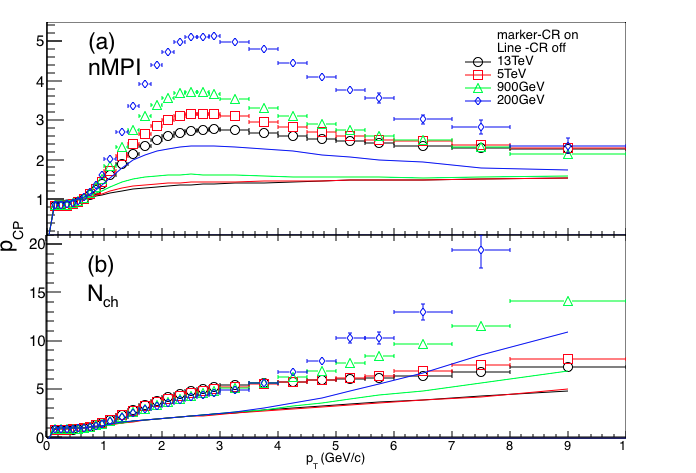}
 \caption{(Color online) $P_{cp}$ as a function of $p_\mathrm{T}$. Top (a) and the bottom (b) panel for the multiplicity classes are computed from from nMPI and $N_{ch}$ (within $|\eta|<0.8$) respectively. Markers represent CR is on, while lines represent CR off. }
 \label{fig:pCP_CROff}
\end{figure}

CMS collaboration has measured finite non-zero elliptic flow $v_2^{sub}\{ 2\}$ in pp collisions at $\sqrt{s}=$ 5.02, 7 and 13 TeV whose magnitude increases with the increase of $\mathrm{N_{trk}^{offline}}$ (multiplicity) suggesting greater collectivity for high-multiplicity events \cite{ppflow}. The notion of collectivity in PYTHIA8 model is incorporated through MPI and CR \cite{pythia6}. In order to see the collective effects or final state effects on $P_{cp}$, we have generated PYTHIA8 events by switching off the CR mechanism. The results thus obtained is compared with the default PYTHIA8 model and is shown in Fig. \ref{fig:pCP_CROff}.

Fig. \ref{fig:pCP_CROff} shows $P_{cp}$ distribution, with and without CR for all energies. In the Fig. \ref{fig:pCP_CROff}(a), we have used nMPI as multiplicity class variable, while in Fig. \ref{fig:pCP_CROff}(b) multiplicity class is calculated from $N_{ch}$ within $|\eta|<0.8$ window. With CR effect, two partons from independent hard scattering at midrapidity can color reconnect and make a large transverse boost. Although the origin of this boost is different in CR compared to hydrodynamics, but this boost effect is similar to how collective phenomena affects hadrons in hydrodynamics \cite{CR-1}. 
Therefore, we do not expect collective behaviour in case of CR off scenario. The $P_{cp}$ in case of CR off in figure \ref{fig:pCP_CROff} is almost flat, and the heavy-ion like $R_{cp}$ behavior is missing.  Results from multiplicity class of $N_{ch}$ shows monotonic increases of $P_{cp}$ for $p_T$ > 4 GeV, with lower magnitude than the CR on scenario.

\section{Summary} \label{summary}
We have presented nuclear modification type phenomena in p+p collisions at $\sqrt{s}=$ 200 GeV, 900 GeV, 5.02 TeV and 13 TeV from PYTHIA8 model. Multiplicity class division is taken from nMPI and $N_{ch}$ variable. Instead of taking percentage of area under nMPI and Nch, we rather used absolute values of these variables to define multiplicity classes. Based on accelerating hydrodynamic description, charge particle multiplicity $dN_{ch}/d\eta >9$ which corresponds to critical energy density, is used to characterise the high multiplicity class.

As expected, we observe multiplicity dependence of $p_{\mathrm{T}}$ spectra for all energies. $p_{\mathrm{T}}$ spectra is seen to be more harder as energy is increased (for high multiplicity classes). When we use multiplicity definition from $N_{ch}$, CR on gives higher  magnitude of $p_{\mathrm{T}}$ spectra than CR-off scenario.
To scale $p_\mathrm{T}$ distribution (both HM and LM class), we have used event average number of pions as this scaling factor shows perfect participant scaling (reported in earlier studies). 

For multiplicity class determination from nMPI, when we use CR on, $P_{CP}$ exhibit an increase from low to intermediate $p_{\mathrm{T}}$, which resembles cold nuclear matter effect observed in heavy ion collisions. $P_{CP}$ shows a decrease from intermediate to high $p_{\mathrm{T}}$, which is seen in heavy-ion collisions as a result of jet quenching. Although the magnitude is higher than unity, which reflects no suppression, but qualitatively the trend is similar to that of heavy-ions. Once we turn off final state effects, such as CR mechanism, $P_{cp}$ is seen to be almost flat and no heavy-ion like effects was observed. So the collectivity like effect in our result is believed to be contributed from CR mechanism. 
However for multiplicity class determination from, $N_{ch}$, we do not see any such heavy-ion like behaviour. We rather see monotonic increase as a function of $p_T$ for all energies. 
For both nMPI and $N_{ch}$, CR on gives higher value than CR off scenario, while both results are higher than unity.

This gives an impression that heavy-ion like effects can also be seen in high multiplicity p+p collisions (with model constraint), we only need to study them differently than the one employed in heavy-ion systems.

\section*{Acknowledgement} 
One of the author SKT, acknowledges support from NKFIH financial grant FK-123842 for this study. Author ANM thanks the Hungarian National Research, Development and Innovation Office (NKFIH) under the contract numbers OTKA K135515, K123815 and NKFIH 2019-2.1.11-T ET-2019-00078, 2019-2.1.11- T ET-2019-00050 and the Wigner GPU Laboratory.


\end{document}